\documentclass[traditabstract]{aa}
\usepackage{graphicx}
\usepackage[varg]{txfonts}
\usepackage{natbib}
\bibpunct{(}{)}{;}{a}{}{,}

\newcommand{\teff}{\ensuremath{T_{\mathrm{eff}}}}
\newcommand{\logg}{\ensuremath{\log g}}
\newcommand{\feh}{\ensuremath{[\rm Fe/\rm H]}}
\newcommand{\afe}{\ensuremath{[\alpha/\rm Fe]}}
\newcommand{\mgfe}{\ensuremath{[\rm Mg/\rm Fe]}}

\newcommand{\Msun}{{\rm M_\odot}}

\newcommand{\intdata}{\textit{iDR1}}

\begin{document}

\title{The Gaia-ESO Survey: radial metallicity gradients and age-metallicity
relation of stars in the Milky Way disk\thanks{Based on observations made with
the ESO/VLT, at Paranal Observatory, under program 188.B-3002 (The Gaia-ESO
Public Spectroscopic Survey).}}
\author{M. Bergemann\inst{\ref{inst1}} \and G. R. Ruchti\inst{\ref{inst2}}
\and
A.
Serenelli\inst{\ref{inst3}} \and S. Feltzing\inst{\ref{inst2}} \and 
A. Alves-Brito\inst{\ref{inst4},\ref{inst25}} \and M. Asplund\inst{\ref{inst4}} \and T.
Bensby\inst{\ref{inst2}} \and P. Gruiters\inst{\ref{inst5}} \and U.
Heiter\inst{\ref{inst5}} \and A. Hourihane\inst{\ref{inst1}}  
\and A.
Korn\inst{\ref{inst5}} \and K. Lind\inst{\ref{inst1}} \and A.
Marino\inst{\ref{inst4}} \and P. Jofre\inst{\ref{inst1}} \and T.
Nordlander\inst{\ref{inst5}} \and N. Ryde\inst{\ref{inst2}} \and 
C.~C. Worley\inst{\ref{inst1}} \and
G. Gilmore\inst{\ref{inst1}} \and S. Randich\inst{\ref{inst6}} \and 
A.~M.~N. Ferguson\inst{\ref{inst10}} \and R.~D.
Jeffries\inst{\ref{inst11}} \and G. Micela\inst{\ref{inst12}} \and I.
Negueruela\inst{\ref{inst13}} \and T. Prusti\inst{\ref{inst14}} \and H-W.
Rix\inst{\ref{inst15}} \and A. Vallenari\inst{\ref{inst16}} \and 
E.~J. Alfaro\inst{\ref{inst22}} \and C. Allende Prieto\inst{\ref{inst7}} \and 
A. Bragaglia\inst{\ref{inst16}} \and S.~E. Koposov\inst{\ref{inst1},
\ref{inst8}} \and
A.~C. Lanzafame\inst{\ref{inst26}}
\and E. Pancino\inst{\ref{inst18}, \ref{inst9}} \and A.
Recio-Blanco\inst{\ref{inst19}} \and R.
Smiljanic\inst{\ref{inst20},\ref{inst21}} \and N. Walton\inst{\ref{inst1}} \and 
M. T. Costado\inst{\ref{inst22}} \and E. Franciosini\inst{\ref{inst6}} \and 
V. Hill\inst{\ref{inst19}} \and C. Lardo \inst{\ref{inst18}} \and P. de
Laverny\inst{\ref{inst19}} \and L. Magrini\inst{\ref{inst6}}
\and E. Maiorca\inst{\ref{inst6}} \and T. Masseron\inst{\ref{inst1}} 
\and L. Morbidelli\inst{\ref{inst6}} \and G. Sacco\inst{\ref{inst6}} \and
G. Kordopatis\inst{\ref{inst1}} \and G. Tautvai\v{s}ien\.{e}\inst{\ref{inst23}}
}

\institute{Institute  of Astronomy, University  of Cambridge,  Madingley Road,
CB3 0HA, Cambridge, UK \email{mbergema@ast.cam.ac.uk} 
\label{inst1} 
\and
Lund Observatory, Department of Astronomy and Theoretical Physics, Box 43,
SE-221 00 Lund, Sweden 
\label{inst2} 
\and
Institute of Space Sciences  (IEEC-CSIC), Campus UAB, Fac. Ci\`encies, Torre C5
parell 2,  08193, Bellaterra, Spain 
\label{inst3} 
\and
Research School of Astronomy \& Astrophysics, Mount Stromlo Observatory, The
Australian National University, ACT 2611, Australia 
\label{inst4} 
\and
Department of Physics and Astronomy, Division of Astronomy and Space Physics,
$\AA$ngstr\"om laboratory, Uppsala University, Box 516, 75120 Uppsala, Sweden
\label{inst5}
\and
INAF - Osservatorio Astrofisico di Arcetri, Largo E. Fermi 5, 50125, Florence,
Italy
\label{inst6}
\and
Instituto de Astrof\'{\i}sica de Canarias, E-38205 La Laguna, Tenerife, Spain
\label{inst7}
\and
Moscow MV Lomonosov State University, Sternberg Astronomical Institute, Moscow
119992, Russia
\label{inst8}
\and
ASI Science Data Center, Via del Politecnico SNC, 00133 Roma, Italy
\label{inst9}
\and
Institute of Astronomy, University of Edinburgh, Blackford Hill, Edinburgh EH9
3HJ, United Kingdom
\label{inst10}
\and
Astrophysics Group, Research Institute for the Environment, Physical Sciences
and Applied Mathematics, Keele University, Keele, Staffordshire ST5 5BG, United
Kingdom
\label{inst11}
\and
INAF - Osservatorio Astronomico di Palermo, Piazza del Parlamento 1, 90134,
Palermo, Italy
\label{inst12}
\and
Departamento de F\'{i}sica, Ingenier\'{i}a de Sistemas y Teor\'{i}a de la
Se$\tilde{\rm n}$al, Universidad de Alicante, Apdo. 99, 03080, Alicante, Spain
\label{inst13}
\and
ESA, ESTEC, Keplerlaan 1, Po Box 299 2200 AG Noordwijk, The Netherlands
\label{inst14}
\and
Max-Planck Institut f\"{u}r Astronomie, K\"{o}nigstuhl 17, 69117 Heidelberg,
Germany
\label{inst15}
\and
INAF - Padova Observatory, Vicolo dell'Osservatorio 5, 35122 Padova, Italy
\label{inst16}
\and
INAF - Osservatorio Astronomico di Bologna, via Ranzani 1, 40127, Bologna, Italy
\label{inst18}
\and
Laboratoire Lagrange (UMR7293), Universit\'e de Nice Sophia Antipolis,
CNRS,Observatoire de la C\^ote d'Azur, CS 34229,F-06304 Nice cedex 4, France
\label{inst19}
\and
Department for Astrophysics, Nicolaus Copernicus Astronomical Center, ul.
Rabia\'{n}ska 8, 87-100 Toru\'{n}, Poland
\label{inst20}
\and
European Southern Observatory, Karl-Schwarzschild-Str. 2, 85748 Garching bei
M\"unchen, Germany
\label{inst21}
\and
Instituto de Astrof\'{i}sica de Andaluc\'{i}a-CSIC, Apdo. 3004, 18080 Granada,
Spain
\label{inst22}
\and
Institute of Theoretical Physics and Astronomy, Vilnius University, Go\v{s}tauto
12, LT-01108 Vilnius, Lithuania
\label{inst23}
\and
Instituto de Fisica, Universidade Federal do Rio Grande do Sul, Av. Bento Goncalves 9500, Porto Alegre, RS, Brazil
\label{inst25}
\and
Dipartimento di Fisica e Astronomia, Sezione Astrofisica, Universit\'{a} di Catania, via S. Sofia 78, 95123, Catania, Italy
\label{inst26}
}

\date{Received date / Accepted date}

\abstract{We study the relationship between age, metallicity, and
$\alpha$-enhancement of FGK stars in the Galactic disk.  The results are based
upon the analysis of high-resolution UVES spectra from the Gaia-ESO large
stellar survey. We explore the limitations of the observed dataset, i.e. the
accuracy of stellar parameters and the selection effects that are
caused by the photometric target preselection. We find that the colour
and magnitude cuts in the survey suppress old metal-rich stars and young
metal-poor stars. This suppression may be as high as 97$\%$ in some regions of
the age-metallicity relationship. The dataset consists of $144$ stars with a
wide range of ages from $0.5$ Gyr to $13.5$ Gyr, Galactocentric distances
from $6$ kpc to $9.5$ kpc, and vertical distances from the plane $0 < |Z| < 1.5$
kpc. On this basis, we find that i) the observed age-metallicity
relation is nearly flat in the range of ages between $0$ Gyr and $8$ Gyr; ii)
at ages older than $9$~Gyr, we see a decrease in [Fe/H] and a clear
absence of metal-rich stars; this cannot be explained by the survey
selection functions; iii) there is a significant scatter of $\feh$
at any age; and iv) $\mgfe$ increases with age, but the dispersion of
$\mgfe$ at ages $>$ 9 Gyr is not as small as advocated by some other studies. In
agreement with earlier work, we find that radial abundance gradients change as a
function of vertical distance from the plane. The $\mgfe$ gradient steepens and
becomes negative. In addition, we show that the inner disk is not only more
$\alpha$-rich compared to the outer disk, but also older, as traced
independently by the ages and Mg abundances of stars.}


\titlerunning{Ages and abundances of stars in the disk}

\authorrunning{M. Bergemann et al.}

\maketitle

%---------------INTRODUCTION--------------------
%
\section{Introduction}

One of the main quests in modern astrophysics is to  understand the formation
history  of the  Milky Way. This requires observational datasets that provide
full phase-space information, ages,  and element abundances for a large number
of stars throughout the Galaxy. The distribution functions of these quantities
present major constraints on the models of the formation and evolution of the
Milky Way \citep[e.g.][]{1992ApJ...391..651B, 1995MNRAS.276..505P, 
1999MNRAS.304..254V, 2000A&A...362..921H, 2001ApJ...554.1044C,
2004ApJ...612..894B, 2005ApJ...630..298B, 2006MNRAS.366..899N,
2009MNRAS.399.1145S, 2013MNRAS.436.1479K, 2013A&A...558A...9M,
2013A&ARv..21...61R}.

In this context, the prime observables are the radial metallicity profiles and
the age-metallicity relation in the Galactic disk
\citep[see e.g. the review by][]{2013NewAR..57...80F}. For a long time, the
mere existence of any age-metallicity relation has been a matter of
great debate. The study by Twarog (1980), which concluded that ages and
metallicities of stars in the disk are uniquely correlated, was the first
important step in putting this discussion onto a modern footing.
\citet{1993A&A...275..101E}, however, was able to show that there is no 
strong correlation. In addition, \citet{2001A&A...377..911F}, and later the
Geneva-Copenhagen Survey \citep{2004A&A...418..989N, 2007A&A...475..519H,
2009A&A...501..941H, 2011yCat..35309138C}, found a large scatter in metallicity
at all ages, and confirmed the tentative existence of an old metal-rich
population. The most recent in-depth analysis of the age-metallicity
relation in the solar neighbourhood is that by \citet{2013arXiv1305.4663H} who
come to similar conclusions.

Combined radial and vertical element abundance profiles provide the
second major observational constraint to the Galaxy formation models.  Attempts
have been made to quantify the metallicity distribution function and abundance
gradients of the disk using data from recent large-scale spectroscopic surveys,
such as SEGUE/SDSS \citep[e.g.][]{2011ApJ...738..187L, 2012ApJ...755..115B,
2012ApJ...753..148B, 2012ApJ...761..160S, 2012ApJ...752...51C}, RAVE
\citep[e.g.][]{2013A&A...559A..59B}, and APOGEE
\citep[e.g.][]{2013arXiv1311.4549A, 2013arXiv1311.4569H}, as well as smaller
samples of stars with high-resolution observations
\citep[e.g.][]{2003A&A...410..527B, 2008MNRAS.384..173F,
2011ApJ...737....9R, 2014A&A...562A..71B}.
However, the bulk of the data has so far come from the traditional techniques,
such as OB stars, Cepheids, and H II regions \citep{2002A&A...381...32A,
2002A&A...384..140A, 2011MNRAS.417..698L, 2013A&A...558A..31L}.
These populations are young, $<1$ Gyr, only providing a snapshot of the present
day abundance pattern. Open clusters and planetary nebulae also provide
information about older populations \citep{2002AJ....124.2693F, 
2010ApJ...714.1096S, 2012AJ....144...95Y, 2014A&A...561A..93H}: open clusters up
to $8-9$ Gyr and planetary nebulae up to $6$ Gyr \citep{2010A&A...512A..19M}.
However, in order to fully study the evolution and build-up of radial and
vertical abundance distributions in the Galactic disk, we need tracers
of stellar populations of all ages, from the earliest epoch of Galaxy formation
to the present day. 

With the data taken with the Gaia-ESO Survey, which form the basis of our work,
 we are in a good position to address these two fundamental problems and set
the stage for the larger datasets coming in the future releases from the survey.
The Gaia-ESO Survey was awarded 300 nights on the Very Large Telescope in
Chile. In the high-resolution ($R = \lambda/\Delta\lambda \sim 47\,000$) mode it
will acquire spectra for about $5\,000$ field stars, probing distances up to 2
kpc from the Sun; $100\,000$ spectra will be acquired in medium-resolution mode
($R \sim 16\,000$), with stars probing distances up to 15 kpc. For the analysis
of the spectra, several state-of-the-art spectrum analysis codes are used
(\citealt{2012Msngr.147...25G}, \citealt{2013Msngr.154...47R}, Smiljanic et al.
in prep). In addition, sophisticated methods based on Bayesian inference are now
available that combine in a  systematic way observational information  and
stellar evolution theory \citep{2004MNRAS.351..487P, 2005A&A...436..127J,
2013arXiv1311.5558S, 2013MNRAS.429.3645S}.

In this study, we use the first results obtained with the Gaia-ESO
survey to study the relationship between age and metallicity in the Galactic
disk. We note that this is an exploratory study and more definitive results will
be available from the next Gaia-ESO data releases with larger samples.
Furthermore, there are other survey papers in preparation that will also
explore the disk and other components in the Milky Way (Recio-Blanco et al.,
Mikolaitis et al., Rojas-Arriagada et al., in prep.). In the
upcoming papers, we will also address other aspects of the Galactic disk
population, including the now-controversial dichotomy of the disk
\citep{2012ApJ...751..131B}, by deriving the kinematics of stars and adding
individual abundances of different chemical elements. We apply the Bayesian
method developed by \citet{2013MNRAS.429.3645S} to derive stellar ages. We
explore the limitations of the observed datasets and perform a detailed
analysis  of the derived stellar properties (mass, ages, metallicity). Finally,
we use these data that span a full range of ages from $1$ Gyr to $14$
Gyr, to study the relationship between elemental abundances and ages of stars,
and discuss the radial abundance profiles in the Galactic disk.  We also discuss
the correlation between Mg abundance and ages of stars. 

The paper is  structured as follows. In Section \ref{sec:obs} observations and
data reduction are described. Section \ref{sec:method} presents the details of
stellar evolution analysis and various tests applied to the input datasets. The
sample selection bias is discussed in Section \ref{sec:bias}. Finally, Section 
\ref{sec:results} describes the results in the context of the Galactic
evolution, with the focus on the age-abundance and spatial distribution of stars
in  the Milky Way  disk, and the conclusions are drawn in Section \ref{sec:conc}
%
%--------------------------GENERAL PROPERTIES---------------------------------
%
\section{Observations and stellar parameters}{\label{sec:obs}}

\begin{figure}[ht]
\centering
\includegraphics[width=0.45\textwidth, angle=0]{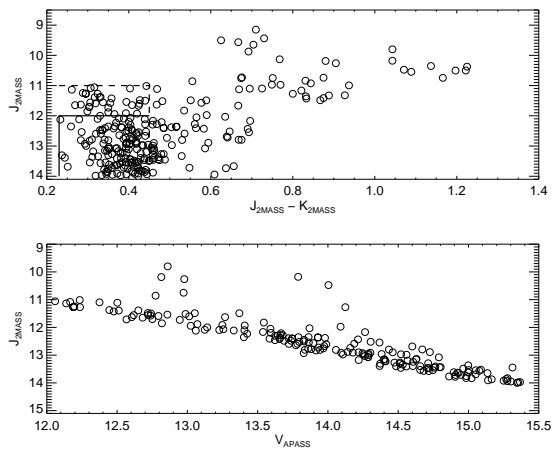}
\caption{The distribution of the observed UVES data sample in the 2MASS (top) 
colour-magnitude plane and as a function of V magnitudes from APASS
(bottom). The photometric box is shown (see text).}
\label{color}
\end{figure}

This work makes use of the  high-resolution UVES
observations\footnote{Resolution $R  \sim 47000$} taken within the first half
year of observations with the Gaia-ESO survey (Jan. 2012  - June 2012), which  
comprises 576 stars. The details on the data reduction will be provided
in Sacco et al. (in prep). Of those, $348$ are field stars while the others are
members of open or globular clusters. The distribution of the observed field
sample in the NIR colour-magnitude diagram and the selection box is shown in
Fig. \ref{color}. The UVES solar neighbourhood targets were chosen according to
their colours to maximise the fraction of un-evolved FG stars within 2 kpc in
the solar neighbourhood. The survey selection box was defined using the 2MASS
photometry:
$12 < J < 14$ and $0.23 < J - K < 0.45 + 0.5 E(B-V)$; if there were not enough
targets, the red edge was extended\footnote{The targets selected before April
(2012) had the brightest cut on J of 11 instead of 12. If the number of
objects in the field within the box was less than the number of UVES fibers,
then the red-edge of the colour-box was shifted to have enough targets to fill
the fibers. In addition, very red objects were selected with a colour-dependent
J cut to prevent low S/N in the optical.}. With these criteria, we are
predominantly selecting FG stars with magnitudes down to $V = 16.5$ (Gilmore 
et al. in prep.).

The stellar parameters, $\teff$, $\log g$, $\feh$, and abundances for the UVES
spectra are determined as follows. The observed  spectra were  processed by $13$
research  groups within the Gaia-ESO survey collaboration with the same model
atmospheres and line lists, but different analysis methods, e.g. full
spectrum chi-square minimisation, using pre-computed grids of model spectra or
calculating synthetic spectra on the fly, or an analysis based on equivalent
widths. The model atmospheres are 1D LTE spherically-symmetric ($\logg  < 3.5$)
 and plane-parallel ($\log g \geq 3.5$) MARCS models
\citep{2008A&A...486..951G}.  The determination of the final set of stellar
parameters based on all those analyses is described in Smiljanic et al. (in
prep). In short, the final parameter homogenisation involves a multi-stage
process, in which both internal and systematic uncertainties of different
datasets are carefully evaluated. Various consistency tests, including the
analysis of stellar clusters, benchmark stars with interferometric and
astroseismic data (\citealt{2013arXiv1309.1099J}, \citealt{2014arXiv1403.3090B},
Heiter et al. in prep.), have been used to assess each group's performance. The
final parameters are taken to be the median of the multiple 
determinations, and the uncertainties of stellar parameters are median absolute
deviations, which reflect the method-to-method dispersion (see Section 5 for
more details). This dataset is available internally to the Gaia-ESO
collaboration, and will be referred to as \intdata. 

\subsection{Comparison with photometry} \label{sec:phot}

The spectroscopic $\teff$ scale can be tested using the Infra-Red Flux Method
\citep[IRFM; ][]{1980A&A....82..249B}. We computed the effective temperature
$T_{J-Ks}$ using the colour - temperature calibrations from 
\citet{2009A&A...497..497G} as described in Ruchti et al. (2011). The $J$, $H$,
and $K_S$ magnitudes were taken from 2MASS and reddening was estimated in a
iterative procedure starting from that found with the
\citet{1998ApJ...500..525S} dust maps. Those stars with $E(B-V) >
0.1$ were then corrected according to the prescription described in Bonifacio et
al. (2000):
$$
E(B-V)_{\rm corrected} = 0.035 + 0.65 * E(B-V)_{\rm Schlegel}.
$$
Finally, we applied the correction for the dust layer, such that the
reddening to a star at distance $D$ and Galactic longitude $b$ is reduced by a
factor $1 - exp(- |~D \sin b~|/ h)$ where $h = 125$~pc. This reduction affects
the stars that are relatively nearby and lie close to the Galactic plane.

\begin{figure}[ht]
\centering
\includegraphics[width=0.4\textwidth, angle=0]{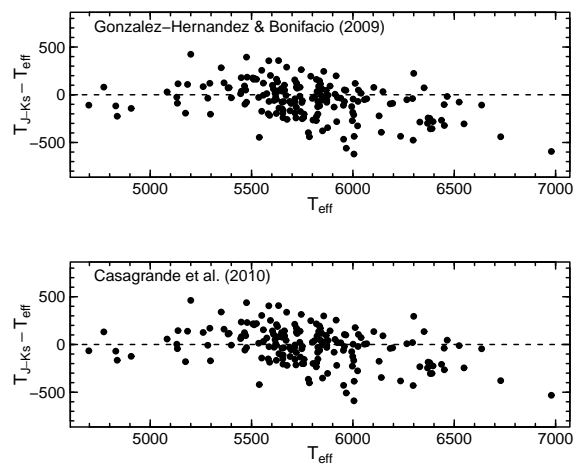}
\caption{Comparison of spectroscopic temperatures and photometric
$T_{J-K}$ temperatures for a subsample of stars (See Section 2.1).}
\label{tphot}
\end{figure}

We compare the results only for a subsample of the stars with
$E(B-V)<0.05$. Fig. \ref{tphot} shows the differences between the spectroscopic
and photometric temperatures for the calibration from
\citet{2009A&A...497..497G} (top panel) and for the calibration by
\cite{Casagrande:2010hj} (bottom panel). The calibrations provide very
similar results; the \cite{Casagrande:2010hj} calibration is $~50$ K warmer. The
both temperature scales appear to be in agreement on average, but there is a
systematic drift at higher $\teff$. The origin of this drift is not
clear: it could be caused by reddening, uncertainties in theoretical photometric
magnitudes, or by spectroscopic uncertainties. For this study, we have decided
to keep all stars. The difference between our scale and that given by IR
photometry is $\Delta \teff = -20 \pm 170$ K 
\cite[calibration by][]{2009A&A...497..497G}. The intrinsic uncertainty of the
$J-K$ IRFM calibration is $150$ K, which may contribute to the wide apparent
spread.
\begin{figure*}[ht!]
\centering
\includegraphics[width=0.8\textwidth, angle=0]{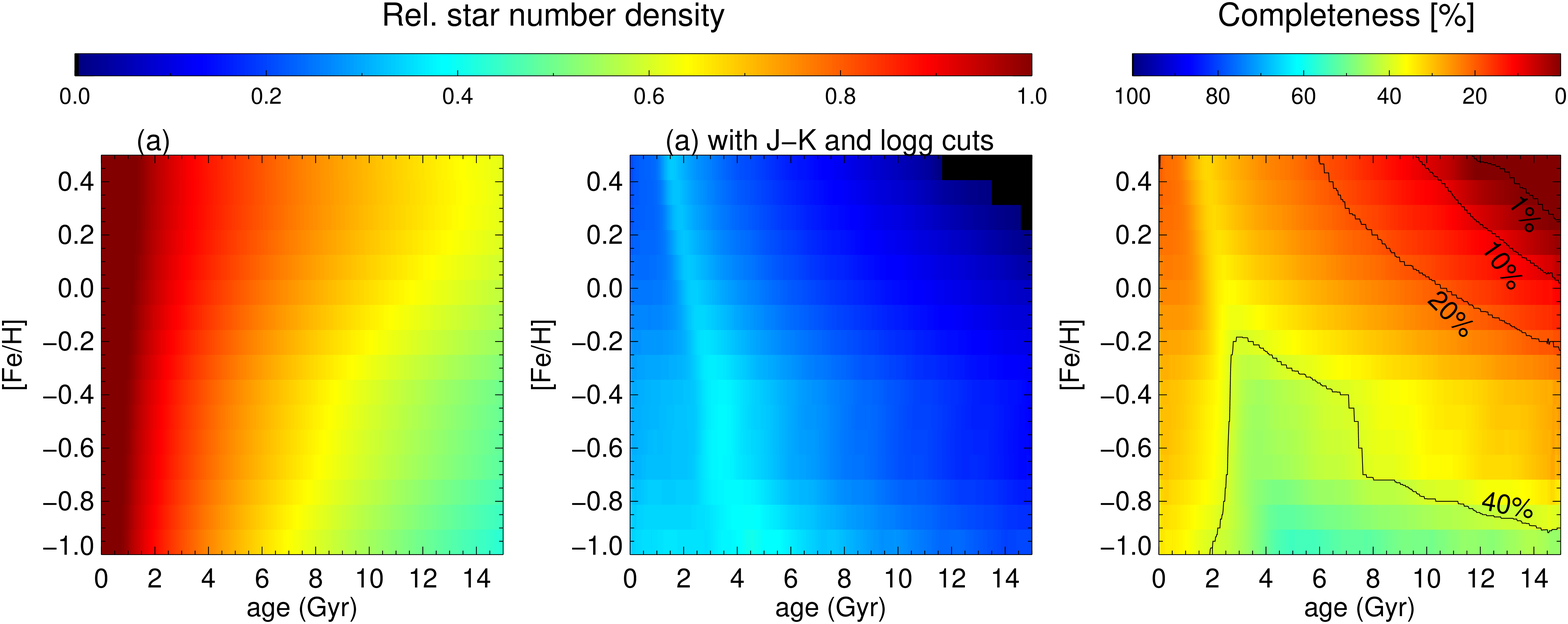}
\includegraphics[width=0.8\textwidth, angle=0]{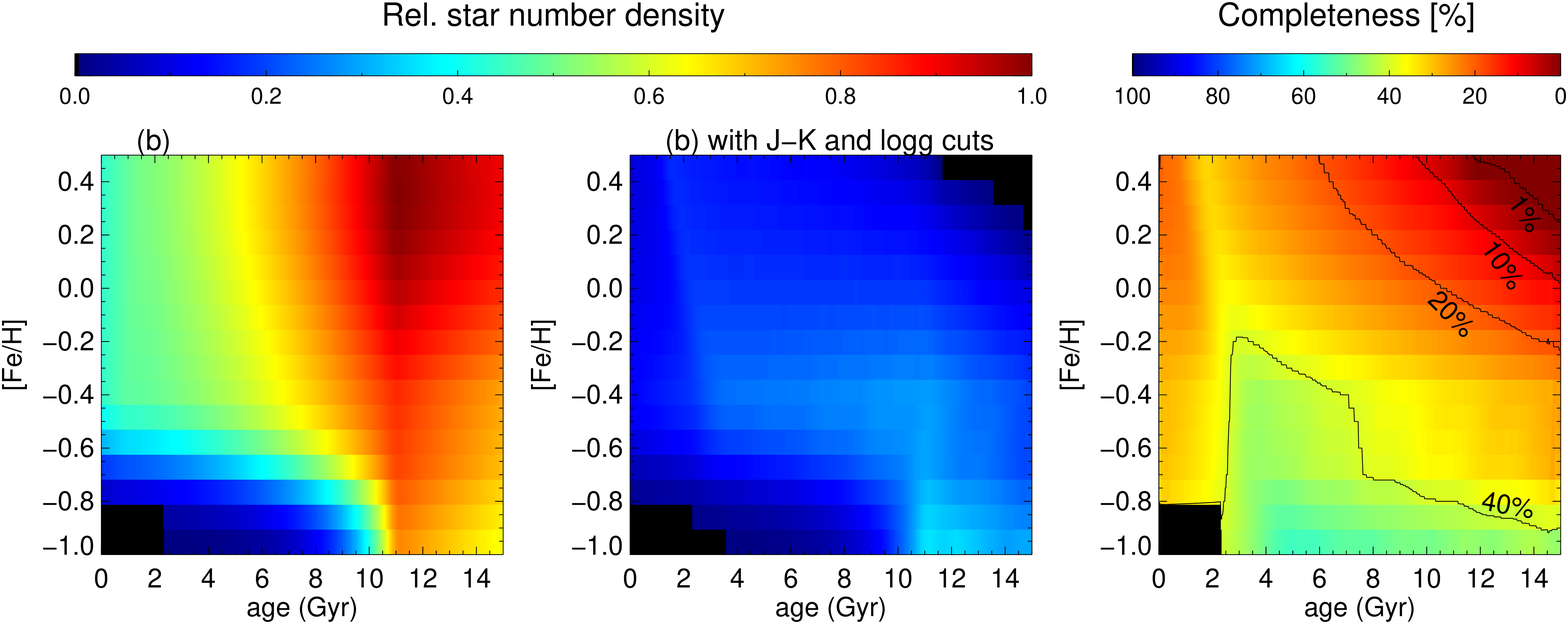}
\caption{Stellar number density computed from the evolution tracks assuming only
IMF (a, top) and IMF+SFR (b, bottom) without colour cuts. Middle panels: with
the colour cut
(0.23 $< J-K <$ 0.45) and stellar parameter limits ($3.5 \leq \logg \leq 4.5$)
to simulate the selection effect of the survey and our stellar sample.
Right: the suppression factor, defined as the ratio of stars with cut to the
number of stars without cut, i.e. the larger the number the more
stars are suppressed (see text).}
\label{colorcut}
\end{figure*}
\subsection{NLTE effects}

One important  problem that needs to be addressed is the influence of the LTE
approximation on the determination of stellar parameters. As shown by
\citet{2012MNRAS.427...27B} and \citet{2012MNRAS.427...50L}, non-local
thermodynamic equilibrium  (NLTE)  effects change  surface  gravities  and
metallicities  of stars, in a way such that  $\log g$ and  $\feh$ obtained from
Fe  lines are systematically higher. However, these studies also showed
that the differences are typically small at metallicity $\feh > -1$.

We have utilised the Spectroscopy Made Easy (SME) spectral synthesis code
\citep{1996A&AS..118..595V} to study the difference between stellar parameters
derived in LTE versus those derived in NLTE. SME has been recently upgraded by
the module to solve for NLTE line formation using the Fe grids described in Lind
et al. (2012).  A complete description of the new module will be presented
elsewhere. Application of the NLTE SME module to the complete \intdata~UVES
dataset analysed in this work revealed that the NLTE effects are minor. For
completeness, we summarise here our results for dwarfs and giants, even
though the latter are not used in the final analysis of the age-metallicity
relation and abundance distributions.

The NLTE excitation equilibrium of Fe\,{\scriptsize I} lines results in somewhat
lower effective temperatures for giants, with stronger effects toward higher
effective temperatures. The effect on surface gravity is more complex. The
direct effect of higher Fe\,{\scriptsize I} abundances results in higher $\log 
g$. At the same time, the lower $\teff$ counteract the direct effect on 
Fe\,{\scriptsize I} abundances, and so $\log g$ comes out with slightly lower
values. The net result is that $\feh$ stays essentially constant. For stars with
$\logg < 3.5$, the mean difference between the NLTE and LTE parameters is:
$-0.25 \pm 17$ K ($\teff$), $-0.05 \pm 0.06$ dex ($\logg$), $-0.01 \pm 0.02$ dex
($\feh$).

For dwarfs, the effect is reversed. Surface gravities and metallicities
increase, although the difference with respect to an LTE analysis is very small.
For stars with $3.5 < \logg < 4.5$, the mean bias in $\logg$ and $\feh$ is $0.03
\pm 0.04$ dex and $0.01 \pm 0.01$ dex, respectively. There is no effect on
$\teff$.

SME is not yet able to compute NLTE abundances for Mg, thus a realistic
assessment of the NLTE effect cannot be done yet. We have checked other
theoretical studies in the literature and found that the bias is also
very small. For example, for the two most important lines of Mg\,{\scriptsize
I}, $5711$ and $5183$ \AA, the difference between the NLTE and LTE abundance is
not larger than $\pm 0.05$ dex for stars with $\logg > 3.5$
\cite[e.g.][]{2000ARep...44..530S}. 

The differences between our LTE and NLTE results are well within the
uncertainties of the stellar parameters and abundances. We thus conclude that
the effect of NLTE in the studied range of stellar parameters will not
significantly bias the resulting age and abundance distributions.
\section{Determination of ages and distances}{\label{sec:method}}
The ages and absolute magnitudes of stars were determined using the Bayesian
pipeline BeSPP\footnote{Bellaterra Stellar Parameter Pipeline} developed by
Serenelli et  al.  (2013).  The grid of  stellar evolutionary tracks  has been
computed with the GARSTEC code \citep{2008Ap&SS.316...99W}. In GARSTEC, nuclear reaction rates are those recommended by \citet{2011RvMP...83..195A},
OPAL opacities by \citet{1996ApJ...464..943I}, complemented at low temperatures
by those from Ferguson et al. (2005), the FreeEOS equation of state from 
\citet{2003ApJ...588..862C}. Convection is treated with the standard mixing
length theory ($\alpha_{\rm MLT}=1.811$, from a solar calibration), and
overshooting has been included as a diffusive mixing process
\citep{1996A&A...313..497F}. More details can be found in Weiss \& Schlattl
(2008) and Serenelli et al. (2013).

Our grid of stellar evolution models spans the mass range $0.6 \leq \rm M/{\rm
M_\odot} \leq 3.5$ with steps of $\Delta M= 0.01{\rm M_\odot}$ and the
metallicity range $-5.0  \leq \feh \leq 0.5$ with steps of $\Delta \feh= 0.1$
for $\feh \leq 0$  and $\Delta \feh= 0.05$ for $\feh > 0$. Models with $\feh
\leq -0.6$ include 0.4~dex $\alpha$-enhancement. A complete description of the
pipeline can be found in \citet{2013MNRAS.429.3645S}. In that paper, we showed
that below  $\logg  \sim 3.5$,   stellar  mass  and  age cannot  be determined
because of the degeneracy  of stellar evolutionary  tracks in  the $\teff-\logg$
plane \citep[see also ][]{1999A&A...352..555A}. Therefore, in this work we
consider only stars with $3.5 < \logg < 4.5$. Due to the $\logg$ cut, $54$
evolved stars are removed from our sample. 

Age was determined by fitting a normal distribution of the probability
distribution function (PDF) in the HR-diagram where the probability density is
higher than 20\% of the  maximum. The best-estimate age is taken to be the
centre of the Gaussian fit and the uncertainties  are determined from the full
PDF as $1-\sigma$ confidence level. We also tested other statistics
using a sample of synthetic stars generated from the tracks (see Appendix A),
but found that neither the weighted mean nor the median of the underlying age
PDF is a satisfactory approximation of the age when the PDF is broad and
asymmetric. Means or medians in truncated PDFs on the young side (or on
the old side) also lead to an overestimation (underestimation) of the true age
of the star.

The BeSPP pipeline has been tested in a variety of ways in Serenelli et  al. 
(2013). However, in this work we are interested in how uncertainties on stellar 
parameters influence the determination of ages. This test was not done in our
previous work, which focused on the relative changes of stellar ages caused
stellar parameters derived under LTE or NLTE. To assess systematic
biases, which could be introduced by the Bayesian method, we performed
test simulations with mock samples of stars drawn randomly from the
evolutionary tracks. We have also checked for the possibility of a systematic
mismatch between $\teff$ and $\logg$ scales from stellar evolution models and
from spectroscopy. From these tests (see Appendix A), we found that the age
estimate is robust and un-biased when its uncertainty is either smaller than
$\sim 30\%$ or smaller than $2~$Gyr. The latter condition is relevant for
younger stars that naturally tend to have larger fractional errors. 
\begin{figure}[ht!]
\centering
\includegraphics[width=0.45\textwidth, angle=0]{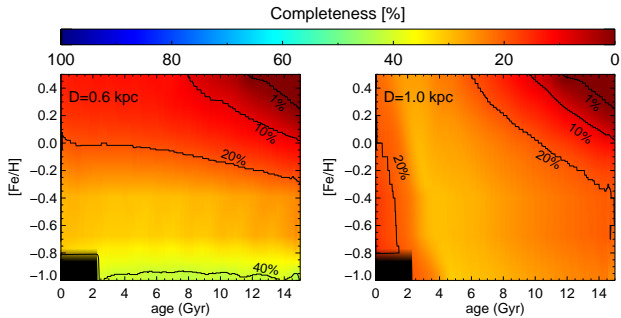}
\caption{Same as Fig. 3 but also including the magnitude cut of the survey: $12
< J < 14$}
\label{magcut}
\end{figure}

Distances were determined following \citet{2011ApJ...737....9R}, using
the absolute magnitudes as described above and the $K_S$ magnitudes from 2MASS.
Reddening is described in Section \ref{sec:phot}. The Galactocentric radial $R$
distance and vertical $Z$ distance from the plane were derived using the
combination of each star's distance from the Sun and Galactic $(\ell, b)$
coordinates. We note that we have assumed that the Sun lies at $R=8.3$~kpc in
the plane. Uncertainties in both $R$ and $Z$ were propagated from the
uncertainty in the distance.
\section{Sample completeness}{\label{sec:bias}}
One of the most difficult but  fundamental tasks is to assess the
completeness of the observed dataset. Various selection
functions (e.g. in colours, magnitudes, or even  $\teff$ and $\logg$) will
reduce the relative number of stars with certain properties. In principle,
accounting for the sample selection is mathematically straightforward (e.g. Rix
\& Bovy 2013); however this has rarely been applied rigorously.

To assess the influence of survey selection cuts, we designed a simple test.
First, we create a stellar population with a Salpeter initial mass
function (IMF), a constant SFR, and a uniform metallicity distribution. The
resulting density of sample stars is shown on the upper-left panel  of
Figure~\ref{colorcut}. The upper-central panel of Figure~\ref{colorcut} shows
the distribution retained after the colour cut used for the UVES sample
is applied, $0.23 < J-K < 0.45$, and with the cuts on surface gravity, $3.5  <
\logg < 4.5$ (see Section~\ref{sec:phot}). The density of stars in the sample is
highest where $J - K = 0.23$, that approximately corresponds to $\sim 6500~$K,
i.e. the blue cut  of the UVES sample. As the figure shows, the cuts
reduce the overall number of stars, with a stronger effect on the oldest
metal-rich stars. By taking the ratio of the stellar number densities computed
with ($N_{\rm cut}$) and without ($N_{\rm no cut}$) the colour and gravity cuts,
we can estimate what fraction of stars is retained in the sample. The sample
completeness is thus defined as:
$$
f = N_{\rm cut}/N_{\rm no cut}
$$
and it is shown in the upper-right  panel of Fig. \ref{colorcut}, along
with the contours of equal probability. The old metal-rich stars, $\feh > 0.2$
can be suppressed almost completely, but also the fraction of old stars with
solar metallicity quickly declines.

We have also considered a more sophisticated model, assuming an SFR varying with
time and with $\feh$ (see Appendix B).  The results of this simulation are shown
in the lower three panels of  Fig.~\ref{colorcut}. The IMF + SFR distribution of
the stellar density (the lower-left panel) is markedly different from the case
of a constant SFR. Because the SFR decreases with time, the stellar number
density always peaks at old ages. In particular, the short e-folding
time at the lowest [Fe/H] is evident by the almost complete absence of young ($<
5$ Gyr) metal-poor stars ($\feh < -0.7$). After applying the colour and gravity
cuts (low-middle panel), the notable difference with respect to the constant SFR
case (top middle panel), in addition to the absence of young metal-poor stars,
is the overall lower density at younger ages for any [Fe/H]. The sample
completeness is shown in the lower-right panel and it is identical to the case
with a flat SFR (upper-right panel). This result can be understood easily.
Stellar number density is constructed by integrating for each age and [Fe/H], 
over the IMF and the SFR. However, the contribution of the SFR to the stellar
number density is only a function of age and  [Fe/H] and therefore cancels out 
when computing the sample completeness as a function of these two quantities. 
Adding a spatial distribution to the discussion also does not affect this
conclusion, as long as it does not include a dependence on the stellar mass.

In the next step, we have imposed the magnitude cut according to the survey
selection. This was achieved by transforming the absolute magnitudes of
synthetic stars to the apparent magnitudes for a range of distances, from $100$
pc to $2$ kpc (typical of the observed sample). The distributions were then
restricted to $J$ magnitudes in the range from $12$ to $14$ mag.  Fig
\ref{magcut} shows two representative cases, where the distance is $0.6$
and $1$ kpc (left and right panel), also including the colour and gravity cuts
as described above. Beyond $0.9$ kpc, the density of young stars with super- and
subsolar metallicity drops, also at distances smaller than $0.6$ kpc metal-rich
stars are suppressed. In either case, the most vulnerable region is the upper
right corner of the age-metallicity plot, that corresponds to old metal-rich
stars.

To summarise, the combined effect of the colour and magnitude limits in the
Gaia-ESO survey catalogue is to under-sample stars that are either young
\textit{and} metal-poor, or old \textit{and} metal-rich. However, the extent to
which these stars are under-sampled, depends quantitatively on their distance
and actual age distribution, which is not known a priori. Furthermore, we have
not accounted for the anisotropy of the inter-stellar reddening, which could
have a differential effect on the typical characteristics of stars observed
within a given photometric box as a function of spatial direction. Therefore, we
see these tests only as a numerical illustration to help us understand
the picture qualitatively, and the results cannot be used to correct the
observed sample.

\section{Results}\label{sec:results}
\subsection{Age-$\afe$ and age-$\feh$ relations}
\begin{figure}[ht]
\centering
\includegraphics[width=0.42\textwidth, angle=0]{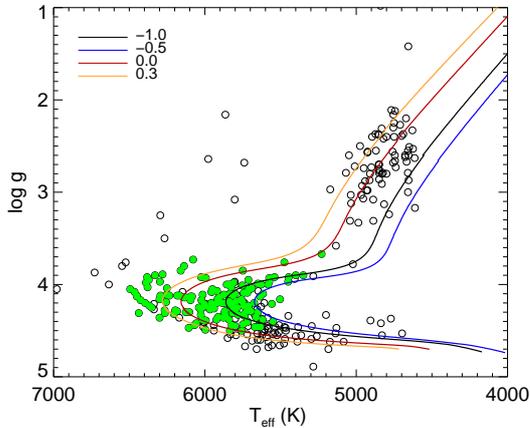}
\caption{The selected UVES sample: (black open circles) - all iDR1 stars;
(filled green circles) - stars that satisfy our selection criteria. 8
Gyr GARSTEC isochrones for different $\feh$ are also shown (see Section 
\ref{sec:method}).
\label{hrd}}
\end{figure}

\begin{figure}[!ht]
\centering
\includegraphics[width=0.5\textwidth, angle=0]{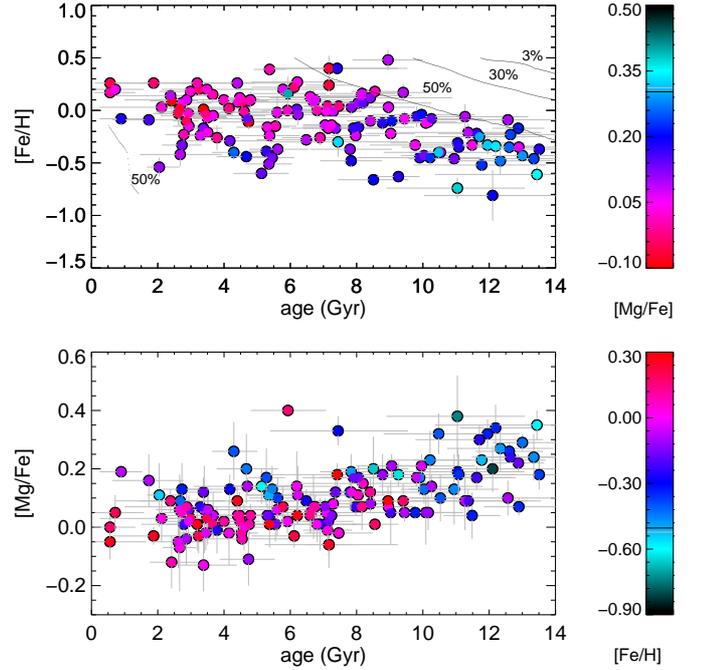}
\caption{Top: Age-metallicity plot for the Milky Way disk. The
contours indicate the relative sample completeness, i.e. the percentage of stars
that would remain in the sample due to the Gaia-ESO survey selection functions,
i.e. IR magnitude and colour cuts, and restrictions imposed on stellar
parameters. Here the magnitude cuts refer to the distance of 1 kpc. For clarity,
this fraction was normalised to its peak value. Bottom: the distribution of
stars in the $\mgfe$ - age plane.}
\label{amr}
\end{figure}

The  main  goal of  this  work  is to  study  the  relation  between ages  and 
abundances of  stars in  the Milky  Way disk. As discussed in Section 
\ref{sec:method},  we apply cuts on $\log  g$ and on age errors. We also remove
stars with uncertainties in Mg abundances larger than $0.15$ dex, which leaves
us with $144$ stars in the final sample (Fig. \ref{hrd}). The mean uncertainties
 are $80$ K in $\teff$, $0.15$ dex in $\logg$, $0.06$ in  $\feh$, and $0.06$ in
$\mgfe$. The results for all selected stars in the age-metallicity and
age-$\mgfe$ planes are shown in Fig. \ref{amr}. The contours in the top panel of
Fig. \ref{amr} indicate the relative sample completeness; the
percentages were normalised to its peak value, which is $\sim 35\%$ 
(Fig.~\ref{colorcut}).

First, we turn to the discussion of the age-metallicity relation (Fig.
\ref{amr}, top panel). The main features demonstrated in earlier observational
studies of the disk by \cite{2001A&A...377..911F, 2013arXiv1305.4663H,
2014A&A...562A..71B} are clearly visible. Between $0$ Gyr and $8$ Gyr
the trend is predominantly flat, and the scatter in $\feh$ is large at any given
age. This is the most populated locus on the diagram, which is unbiased
according to our tests. After $8$ Gyr, there is a clear decline in $\feh$, such
that the older stars are progressively more metal-poor. Our result does not
support the analysis by \citet{2011A&A...530A.138C}, which is based on the
photometric metallicities and ages of stars in the Geneva-Copenhagen Survey. The
authors find no age-metallicity relation; the stars are homogeneously
distributed in metallicity in any age bin up to 12 Gyr \citep[][their Fig.
16]{2011A&A...530A.138C}. While qualitatively, the mean metallicity of the
sample for old ages could be affected by our sampling bias against old and
metal-rich stars, Fig. 6 shows that the suppression relative to the most
populated part of the plot is not larger than $50 - 70\%$. The fact that no
metal-rich star is observed with age $> 10$ Gyr may indicate  that such stars
are rare, if they exist at all in the solar neighbourhood.

Figure \ref{amr} (bottom panel) shows $\mgfe$ ratios as a function of
age, colour-coded with $\feh$. The oldest stars with ages $> 12$ Gyr show
$\mgfe$, from $0$ to $0.4$ dex, and a broad range of metallicity, from solar to
$\feh \sim -1$. There is little evidence that the relation tightens at ages
greater than $9$ Gyr in our sample, as advocated e.g. by
\citet{2013arXiv1305.4663H} who used a subsample of 363 stars from the
\citet{2012A&A...545A..32A} sample of 1111 FGK stars.
\citet{2014A&A...562A..71B} also found a knee at $9$ Gyr, with a clear increase
in $\mgfe$ with age (their Fig. 21), albeit with a notably larger scatter at
ages $> 11$ Gyr than in \citet{2013arXiv1305.4663H}. It is possible that the
larger scatter in our sample at old ages is due to the fact that we include
stars with relatively large age uncertainties. However, from our analysis of the
selection effects, it is to be expected that some fraction of $\alpha$-poor old
stars could be artificially suppressed for the same reasons as discussed above,
akin to the $\feh$ suppression shown in the age-metallicity plot.  Regardless of
these effects, the trends at old age, as seen by \citet{2013arXiv1305.4663H} and
\citet{2014A&A...562A..71B} fit within our $\mgfe$-age relation.

\begin{figure}[ht!]
\centering
\includegraphics[width=0.45\textwidth, angle=0]{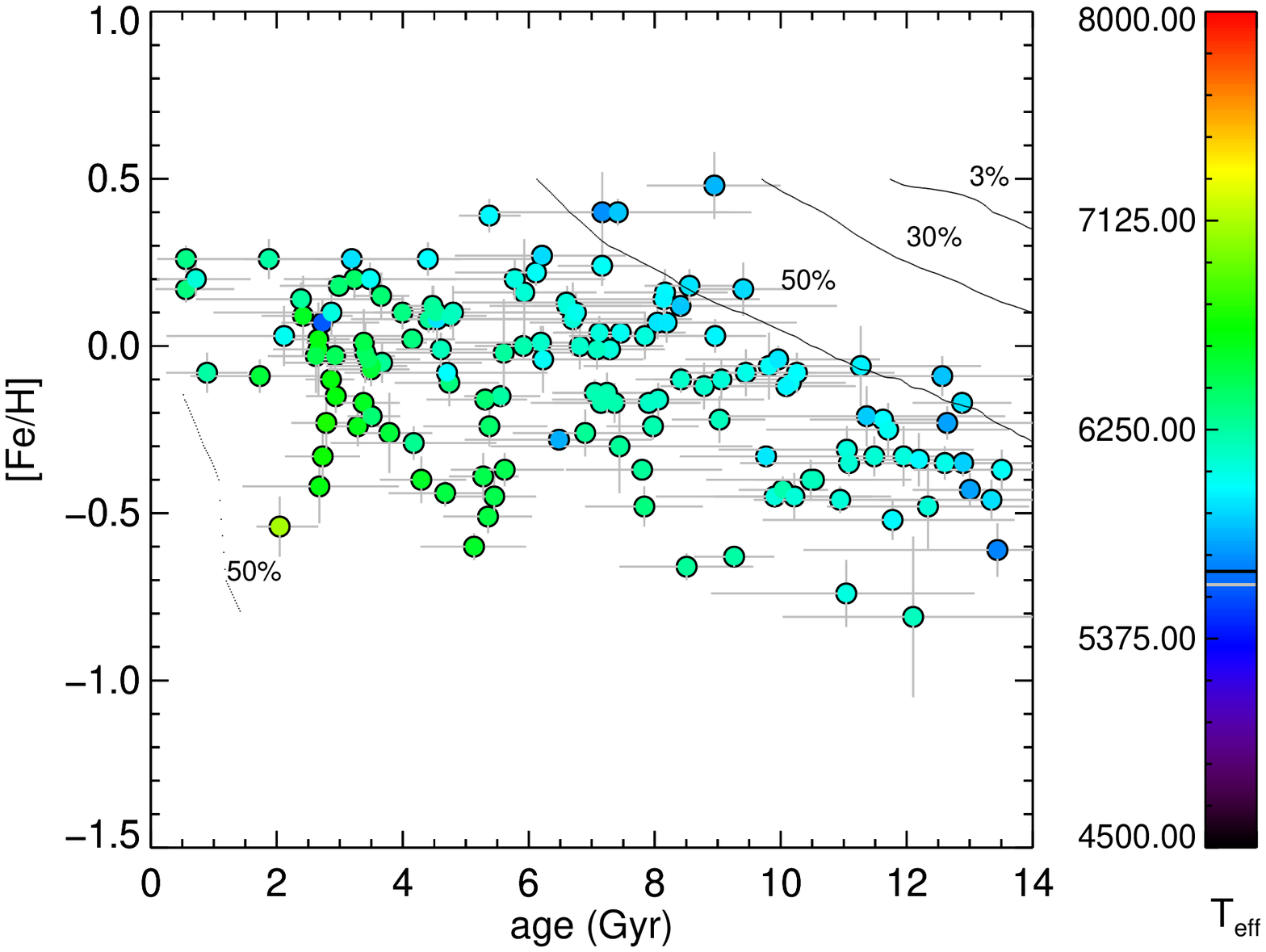}
\includegraphics[width=0.45\textwidth, angle=0]{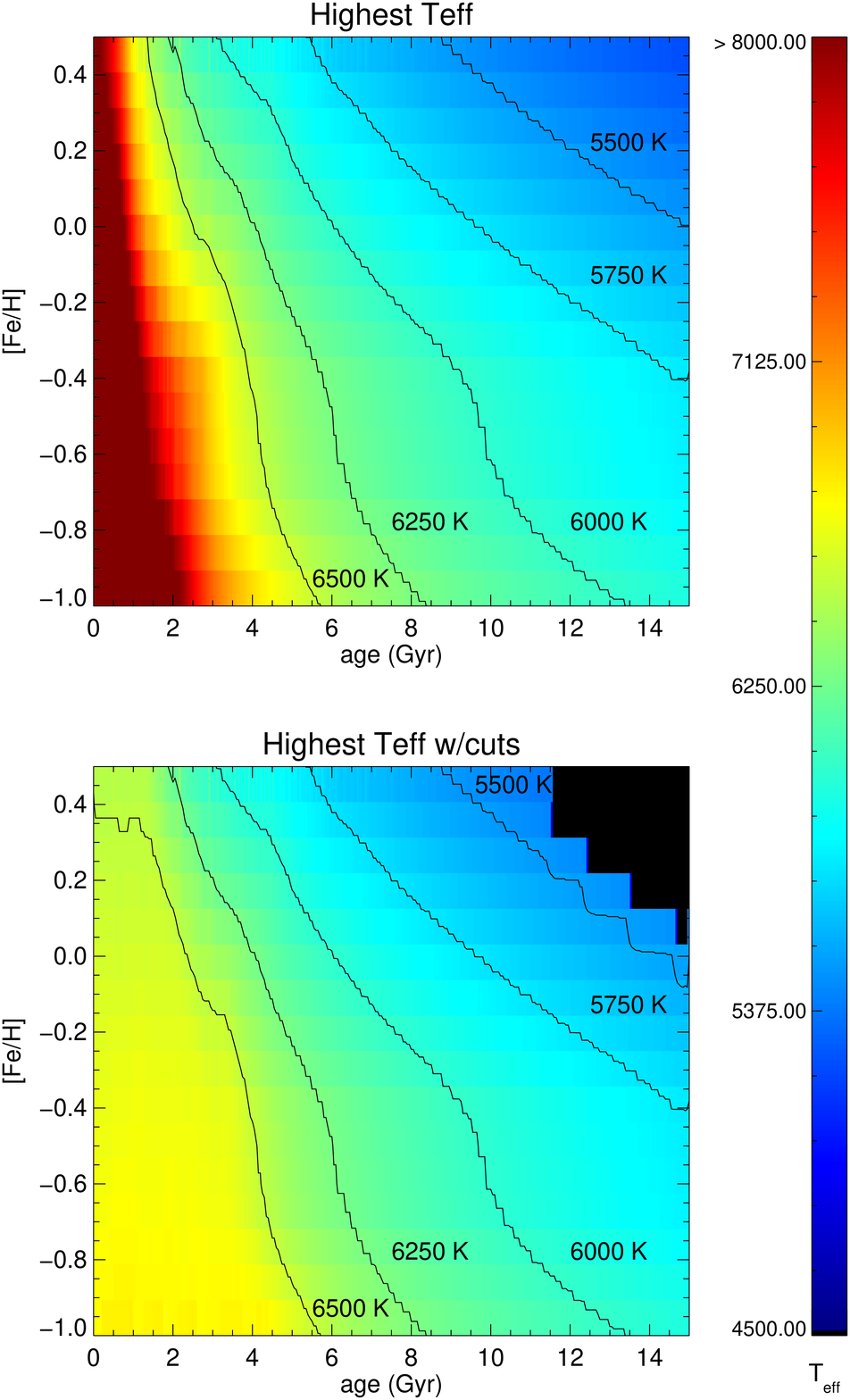}
\caption{Top panel: the observed stars in the age-$\feh$ plane,
colour-coded with their $\teff$. Bottom two panels: The highest possible $\teff$
for a given age and $\feh$ obtained from the  stellar evolution tracks without
(middle) and with (bottom) photometry and $\log g$ cuts. Selected curves of
constant $\teff$ are given for reference. For $\teff \le 6500~\rm{K}$, the
highest 'observable' values of $\teff$ do not depend on the cuts imposed on the
sample and are simply the result of stellar evolution effects.}
\label{amr1}
\end{figure}

Finally, one interesting feature of the age-metallicity relation
deserves a comment. Fig.~\ref{amr1} (top panel) shows the observed stars in the
age-metallicity plane colour-coded with their $\teff$. The obvious correlation
with effective temperature is striking yet it can be easily explained based on
the similar considerations as in Section \ref{sec:bias}. In the middle and
bottom panel of Fig. \ref{amr1}, we also show the maximum $\teff$ to be expected
in the $\feh$ - age plane based on stellar evolution models, without (middle)
and with (bottom) cuts on the photometry and stellar parameters. The colour and
$\log g$ cuts affect the bottom left corner of the age-metallicity plot. Young
hot stars are suppressed because of the blue colour cut in the Gaia-ESO UVES
sample selection. However, from the comparison of these two plots we see that
the trend for $\teff \le 6500~\rm{K}$ does not depend on selection effects. The
segregation of the sample according to the $\teff$ is thus simply the
consequence of stellar evolution: the highest 'observable' temperature of stars 
decreases with age and metallicity, and naturally leads to the $\teff$
distribution shown in Fig.~\ref{amr1}. Such a trend must be present in any
survey, as long as the probed $\teff$ range is not too narrow.
\subsection{Abundance gradients as a function of radial and vertical distance}
\begin{figure*}[!ht]
\centering
\includegraphics[width=0.8\textwidth, angle=0]{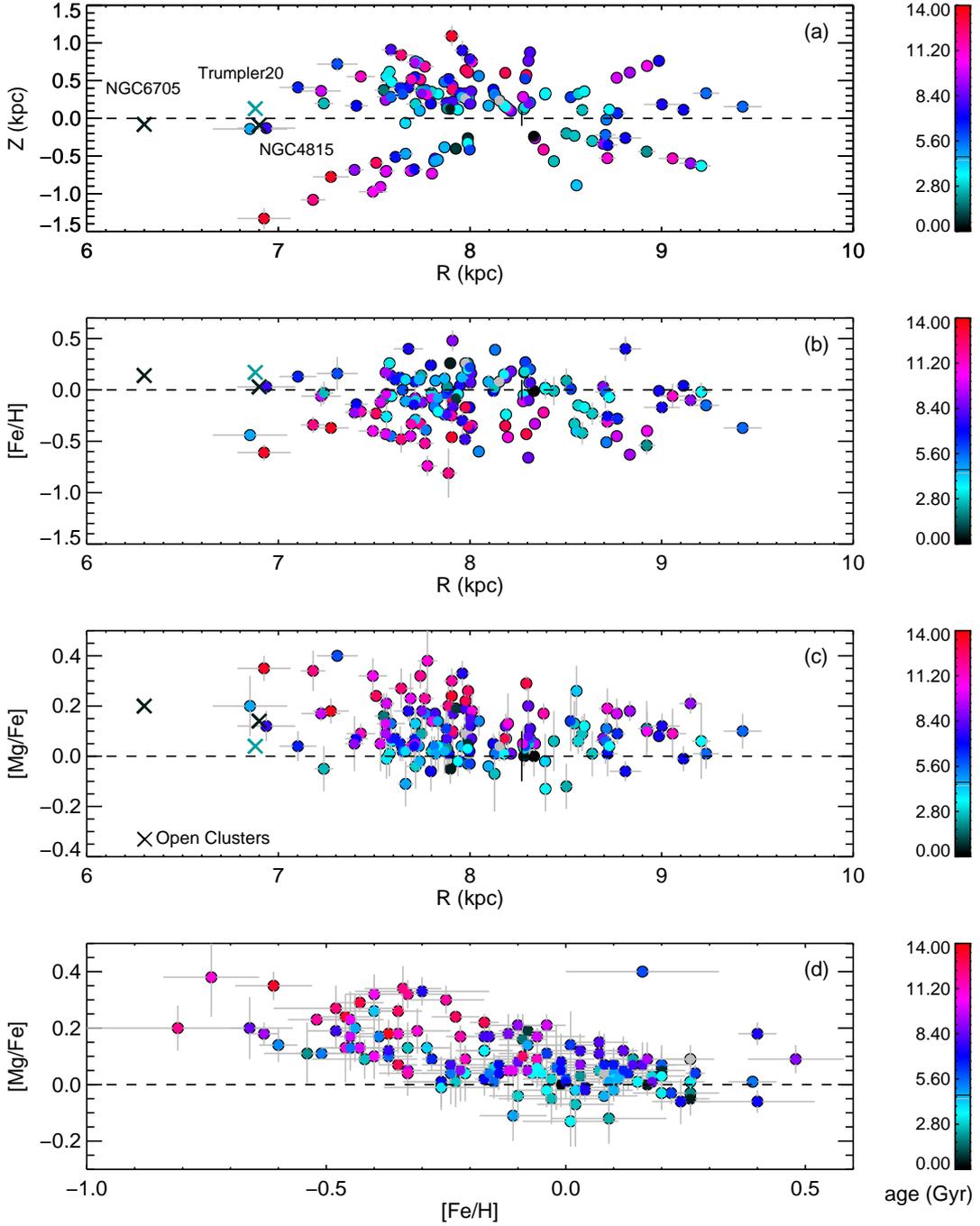}
\caption{The final stellar sample. Top to bottom: a) the distribution of stars
in the $Z$ vs $R$ plane; b and c) radial gradients in the MW disk for [Fe/H] and
[Mg/Fe]; d) the $\feh-\mgfe$ relation. Colour represents age of a star. The
three young open clusters, which are available in UVES iDR1, are shown with
large crosses.
\label{lowhighs}}
\end{figure*}
Figure \ref{lowhighs} shows how stars are distributed with the vertical
distance from the plane $|Z|$ and Galactocentric distance $R$, colour-coded
according to their ages. The middle panels show the gradients of $\feh$ and
$\mgfe$ as a function of $R$; the bottom panel is the $\mgfe$-$\feh$ relation.
The three open clusters that are available in the UVES iDR1, NGC 6705, Trumpler
20, and NGC 4815 \citep{2013arXiv1312.6472M} are also plotted.

There are obvious differences in the age, metallicity, and Mg gradients
with $R$. The younger stars with ages $\leq 7$ Gyr show an outward decrease in
the mean $\feh$, slightly stronger compared to the previous studies of young
disk populations, i.e Cepheids, OB stars, and H II regions (Andrievsky et al.
2002a,b, \citealt{2013A&A...558A..31L}). Using only stars close to the
plane, within $|Z|$ of $300$ pc, we derive $\Delta \feh/\Delta R = -0.068 \pm
0.014$ dex$/$kpc or $\Delta \feh/\Delta R = -0.076 \pm 0.010$ dex$/$kpc (stars
with ages $\leq 7$ Gyr). Analysing RAVE spectra of disk dwarfs,
\citet{2013A&A...559A..59B} found that $\Delta \feh/\Delta R = -0.065$ dex$/$kpc
for $Z_{\rm max} < 400$ pc. For $300 < Z_{\rm max} < 800$ pc, we obtain
$\Delta \feh/\Delta R = -0.114 \pm 0.009$ dex$/$kpc. The steepening of the
gradient is surprising given the other studies found flatter or even positive
radial metallicity gradients at larger vertical distances from the plane
(\citealt[][]{2006ApJ...636..804A},\citealt[][]{2013arXiv1311.4569H,
2013arXiv1311.4549A}, \citealt[][]{2011A&A...525A..90K}), but this could reflect
our small dataset or the fact that the studies probe much larger $|Z|$.

The old stars in our sample tend to have lower $\feh$ even in the inner
disk (Fig. \ref{lowhighs}, (b)), and our $\feh$ gradient for old stars with ages
above $12$ Gyr is $\Delta \feh/\Delta R = 0.069 \pm 0.044$, 
positive but with a larger uncertainty and potentially consistent with being
flat. However, we also note that this apparent age dependence of the radial
metallicity gradient is also related to the fact that we observe older stars at
larger distances from the plane. As seen in the top (a) panel of Fig.
\ref{lowhighs}, the majority of old stars are also located at $|Z| > 800$ pc and
our sample does not have directional isotropy: we probe larger distances in the
inner disk. Therefore, we cannot yet make more definitive statements about the
gradients for old stellar population, or at high $|Z|$.
\citet{2004A&A...418..989N} noted a change of slope of the radial [Fe/H]
gradient in the disk. They find a progressively flattening $\delta \feh/\delta
R_{\rm mean}$ with increasing age, which even becomes positive for ages $>10$
Gyr\footnote{Nordstrom et al. (2004) studied gradients as a function
of the mean radius R$_{\rm mean}$}. It is likely that their results also reflect
the effect of observing older stars at larger $|Z|$ from the plane. 

The radial [Mg/Fe] gradient is shown in the panel (c) of Fig. \ref{lowhighs}
colour-coded with age. For the stars with $|Z| \leq 300$ pc, the radial gradient
in the sample mean of the $\mgfe$ abundance is close to
zero, $\Delta \mgfe/\Delta R = 0.021 \pm 0.014$. However, the gradient
becomes negative and steepens with increasing distance from the plane. For the
stars with $0.3 < |Z| < 0.8$ kpc in our sample, the most populated bin in our
sample, we infer $\Delta \mgfe/\Delta R = -0.045 \pm 0.011$. It appears that the
gradient steepens further at even larger vertical distance, $|Z| > 0.8$ kpc.
Moreover, we find that the gradient of $\mgfe$ becomes negative and very
steep for older stars, i.e. $\Delta \mgfe/\Delta R = 0.015 \pm 0.014$ (ages
$\leq 7$ Gyr) and $\Delta \mgfe/\Delta R = -0.071 \pm 0.029$ (ages $\geq 12$
Gyr). In words, the outer disk has \textit{less high-Mg and old stars} than the
inner disk. These results are very interesting, and they support and extend the
results of the earlier studies, including \citet{2010A&A...516L..13B,
2011ApJ...735L..46B} and \citet{2012ApJ...753..148B}. The former focussed on the
inner ($4 < R < 7$ kpc) and outer ($9 < R < 13$ kpc) disk giants. Their main
conclusion is that while the inner disk giants chemically behave as the solar
neighbourhood stars in terms of the "two" disk components (low $\afe$ thin-disk
stars vs high $\afe$ thick-disk stars), the outer disk is only formed of
"thin-disk"-like stars. \citet{2012ApJ...753..148B} found that the incidence of
$\alpha$-enhanced stars falls off strongly towards the outer disk using G-dwarfs
from the SEGUE survey. In this work, we are also able to add, for the first
time, ages to the inner-outer disk  picture. The scatter of $\feh$, $\mgfe$, and
ages of stars  towards the inner disk is \textit{larger than in the outer disk}.
This means that the chemical evolution models for the Galactic disk which
predict smaller abundance scatter in the inner disk \citep{2000A&A...362..921H}
are disfavoured by our results.

The bottom (d) panel of Fig. \ref{lowhighs} shows the $\mgfe$-$\feh$
relation. The results are not un-expected: the solar-metallicity stars are
predominantly young. At lower metallicity, we see younger stars with low Mg/Fe
abundance ratios, but also very old Mg-enhanced stars. Whether these are two
different components, each with its own history, or a continuous distribution of
stars cannot be firmly established yet.

In general, our results may have several interpretations \citep[see][
for a review of disk formation scenarios]{2013NewAR..57...80F}. Some
observational studies \cite[e.g.][]{2008ApJ...673..864J} decompose the disk into
the thin and thick components, which formed in different episodes of Galaxy
formation. Models which assume different evolutionary history of the two disks
\cite[e.g.][]{1997ApJ...477..765C, 2001ApJ...554.1044C} are broadly consistent
with observations. Stars in the thin disk are young $<8$~Gyr and $\alpha$-poor,
while the thick disk is older, metal-poor and $\alpha$-enhanced
\citep{2003A&A...410..527B,2011A&A...535A.107K,2013MNRAS.436.3231K}. In this
interpretation, at the smallest distances from the plane, our age-metallicity
relation is dominated by the thin disk. In the intermediate vertical distance
bin, our data would indicate a presence of both the thin and thick disk. 

While the discrete nature of the disk is one possibility, our data can
be also explained by other Galaxy formation models, which do not explicitly
divide the disk into subcomponents \citep{2000A&A...362..921H,
2008ApJ...684L..79R, 2013arXiv1308.2061R, 2013MNRAS.436.1479K,
2013A&A...558A...9M, 2014arXiv1401.5796M}. For example, the
semi-analytical model by \citep[][see their Fig. 6]{2009MNRAS.399.1145S}, which
assumes radial migration driven by transient spiral arms, is consistent with our
data both what concerns the age-metallicity relation shape and the $\feh$
scatter at a given age. In accord with our results, this model also predicts
the apparent bimodality in the \feh-\afe plane, just as we observe (Fig.
8, bottom panel). This effect is not a consequence of a star formation hiatus
\citep[as e.g. in][]{2001ApJ...554.1044C}, but the $1~$Gyr timescale of SN Ia
coupled with secular heating and churning \citep[][a simple analytical
explanation is given in their Appendix B]{2009MNRAS.396..203S}. There exist
other types of models that also show consistencies with our data. For
example, hydrodynamical N-body simulations \citep{2012MNRAS.426..690B,
2013A&A...558A...9M} form a thick-disk like component through early gas-rich
mergers and radial migration driven by external and internal perturbations. The
most recent simulations by  \citet[][their Fig. 9]{2014arXiv1401.5796M} predict
that mono-age populations have radial abundance gradients that do not change
with vertical distance from the plane. The change of radial gradients with $|Z|$
is thus caused by a different mix of stellar ages at different altitudes. We
cannot yet test this hypothesis because of the sparse sampling of stars.
However, we confirm that the positive gradient in $\mgfe$ at low $|Z|$ changes
to negative at larger vertical distance from the plane, we also observe a
similar structure in the mono-age bins. To put more quantitative constraints on
their prediction, more data is needed. Therefore we leave this analysis for the
next paper, which will include a much larger number of stars from the Gaia-ESO
DR2 dataset.
%
%--------------------------------------------------------------
%
\section{Summary}\label{sec:conc}

The main goal of our work is to study the relationship between age and
metallicity and spatial distribution of stars in the Galactic disk from
observational perspective.

The stellar sample includes several hundreds of stars observed at
high-resolution (with the UVES instrument at VLT) within the first six months
of the Gaia-ESO survey. Stellar parameters and abundances were determined with
1D local thermodynamic equilibrium (LTE) MARCS model atmospheres, by combining
the results obtained by several different spectroscopic techniques. We then
derived the mass and age of each star, as well as its distance from the Sun,
using the Bayesian method developed in Serenelli et al. (2013). The results were
validated against astero-seismology and the IRFM method.  We took special care
to quantify the survey selection effects in the interpretation of the final
parameter distributions. For this, we performed a series of tests using a mock
stellar sample generated from theoretical stellar models, which was  subject to
various cuts in the space of observables. The outcome of these tests allowed us
to define the optimal dataset to address the science problems. In the final
sample, we have $144$ stars with a wide range of ages ($0.5 - 13.5$ Gyr),
Galactocentric distances from $6$ kpc to $9$ kpc, and vertical distances from
the plane $0 < |Z| < 1.5$ kpc. The sample is not large, however it should be 
noted that the study is exploratory, i.e. laying out the method and the
properties of the observed field sample of stars in the Gaia-ESO Survey.  At the
end of the survey we expect to have a one order of magnitude increase in the
stellar sample.

Our findings can be summarised as follows.

First, the observed age-metallicity relation is fairly flat in
range of ages between $0$ and $8$ Gyr (or equivalently in the young disk, for
stars close to the plane). We also confirm the previous results
that there is a significant scatter of metallicity at any age. A steep decline
in [Fe/H] is seen for stars with ages above $9$ Gyr. The colour and
magnitude cuts on the survey suppress very old metal-rich stars and young
metal-poor stars. However, the suppression relative to the most populated locus
on the age-metallicity relation is not larger than $50 - 70\%$. In our sample,
no solar-metallicity star is observed with age $> 10$ Gyr. This may indicate
that such stars are rare, if they exist at all in the solar neighbourhood.

$\mgfe$ ratios correlate with age, such that $\alpha$-rich stars are
older, but the dispersion of $\mgfe$ abundances is not small at any age. In
particular, the old stars with ages above $9$ Gyr have a range of
$\alpha$-enhancement, from $0$ to $0.4$ dex, and metallicity from solar to $\feh
\sim -1$. This contrasts with the very tight correlation of $\mgfe$ and ages for
old stars as suggested by Haywood et al. (2013). However, the trends found by
Haywood et al. (2013) and Bensby et al. (2014) overlap with our $\mgfe$-age
relation.

We also show, in agreement with earlier observational and theoretical work, that
the radial abundance gradients (Fe, Mg) change as a function of vertical
distance from the plane, or the mean age of a population. At $|Z| \leq
300$ pc, we find $\Delta \feh/\Delta R = -0.068 \pm 0.014$ and $\Delta
\mgfe/\Delta R = 0.021 \pm 0.014$. For the most populated $|Z|$ bin in our
sample, $300 \leq |Z| \leq 800$ pc we infer $\Delta \mgfe/\Delta R = -0.045 \pm
0.011$, and $\Delta \feh/\Delta R = -0.114 \pm 0.009$. The picture is
not too different when separating stars according to their age: the gradient of
$\mgfe$ is close to zero for younger stars, but becomes negative and very steep
for the older stellar population, i.e. $\Delta \mgfe/\Delta R = 0.015 \pm 0.014$
(ages $\leq 7$ Gyr) and $\Delta \mgfe/\Delta R = -0.071 \pm 0.029$ (ages $\geq
12$ Gyr). The anisotropic distribution of stars in our sample, i.e. larger
fraction of stars observed in the inner disk, should be kept in mind when
interpreting these gradients.

To summarise, perhaps the most important result of this analysis is how the
properties of the dominant stellar populations in the disk change as we probe
stars with different ages, distances, Mg abundances, and metallicities. We find
more older, $\alpha$-rich, and metal-poor stars in the inner disk. In the outer
disk, stars are on average younger and $\alpha$-poor. Although our current
sample is small, our results lend support to current pictures of the formation
of the Galactic disk, such as the inside-out formation. This initial dataset
illustrates the potential of the Gaia-ESO Survey. With upcoming data
releases, we will have a large database of spectra that have been reduced and
analysed in a systematic and homogeneous manner. We will thus be able to
distinguish among the many theories of the formation of the disk of the Milky
Way. 
%
%
%--------------------------------------------------------------
%\bibliographystyle{aa}
\bibliography{references}
%--------------------------------------------------------------
%
\begin{acknowledgements}
This work was partly supported by the European Union FP7 programme through ERC
grant number 320360. AS is supported  by the MICINN grant AYA2011-24704 and by
the ESF EUROCORES Program EuroGENESIS (MICINN grant EUI2009-04170). The results
presented here benefited from discussions held during Gaia-ESO workshops and
conferences supported by the ESF (European Science Foundation) through the GREAT
(Gaia Research for European Astronomy Training) Research Network Program. Based
on data products from observations made with ESO Telescopes at the La Silla
Paranal Observatory under programme ID 188.B-3002. This work was partly
supported by the European Union FP7 programme through ERC grant number 320360
and by the Leverhulme Trust through grant RPG-2012-541. We acknowledge the
support from INAF and Ministero dell' Istruzione, dell' Universit\`a' e della
Ricerca (MIUR) in the form of the grant "Premiale VLT 2012". GRR and SF
acknowledge support by grant No. 2011-5042 from the Swedish Research Council. We
acknowledge support from the Swedish National Space Board (Rymdstyrelsen). T.B. was funded by grant No. 621-2009-3911 from The Swedish Research Council. 

\end{acknowledgements}

\begin{appendix}

\section{Tests}

To check for systematic biases caused by the method itself, we performed 
Monte-Carlo  simulations by generating  a random  sample  of synthetic  stars
assuming a Salpeter IMF, a constant  star formation rate, and a uniform $\feh$
distribution over  time. The sample is  restricted to the  low-mass range $0.7
\leq \rm M/{\rm M_\odot}  \leq 1.5$ and $-2.2 \leq  \feh \leq 0.4$,
representative of the Gaia-ESO survey UVES sample. Mass and  $\feh$ are
discretised to match the values present in the  grid. Along each track, $\teff$ 
and $\logg$ are interpolated to the  age of  the synthetic  star.  The synthetic
 stars were then assigned individual  uncertainty values, comprising two 
sources. The random uncertainty component was  drawn from the empirical 
probability distributions
constructed from  the uncertainty  distribution  in the  actual  UVES sample. In
addition,  we accounted  for the  possibility  of a  systematic  error. The
central values of the simulated data were perturbed by introducing noise drawn
from  uniform  distributions in  the  range  (-100,100)~K, (-0.1,0.1)~dex  and
(-0.05,0.05)~dex for $\teff$, $\logg$, and $\feh$ respectively. The tests showed
that the age estimate is robust if the uncertainty defined as described  above 
is smaller  than  $\sim 30\%$  or  than  2~Gyr. The latter condition is relevant
for younger stars that naturally have larger fractional errors. 
\begin{figure}[ht]
\centering
\includegraphics[width=0.5\textwidth, angle=0]{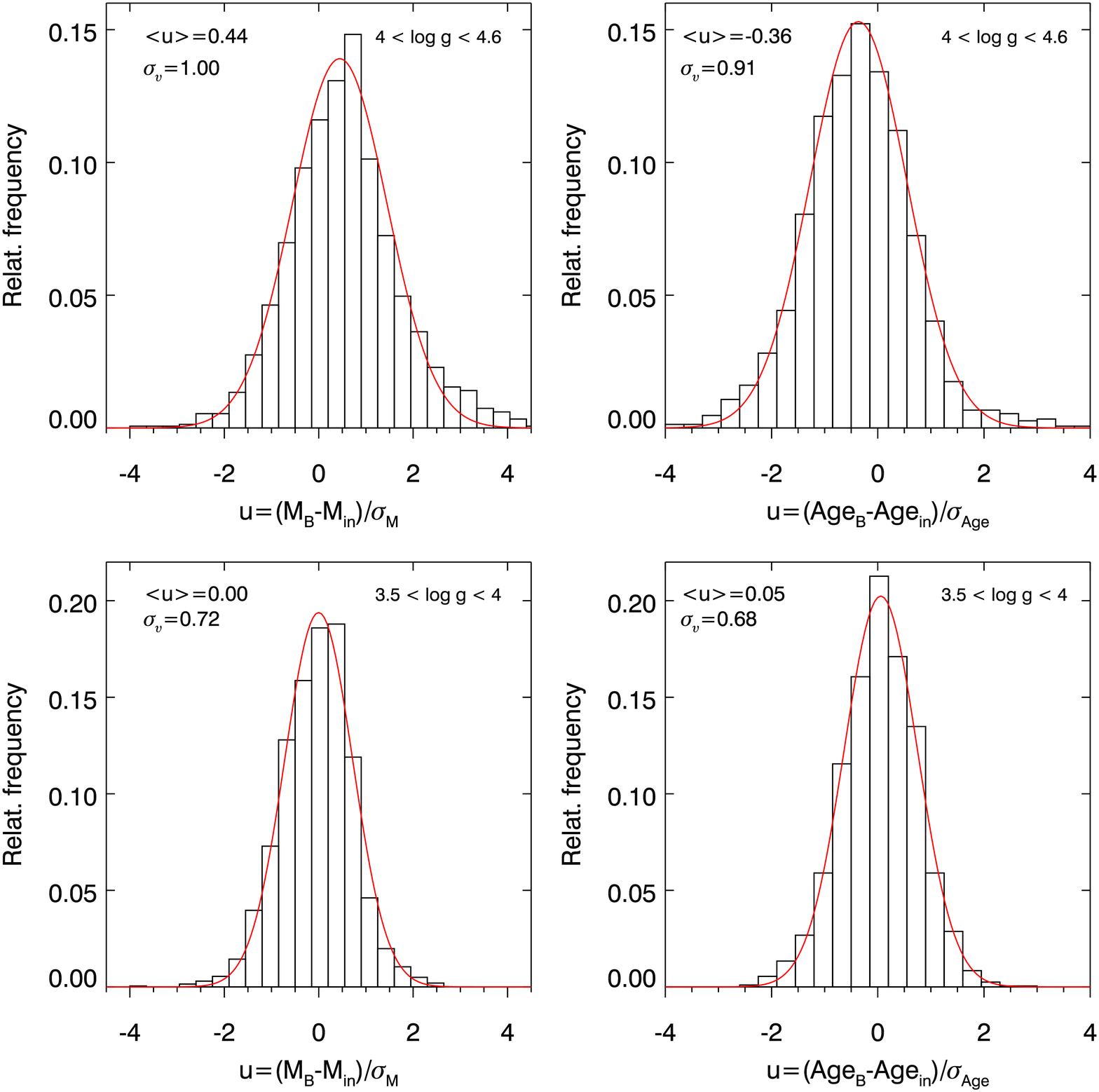}
\includegraphics[width=0.5\textwidth, angle=0]{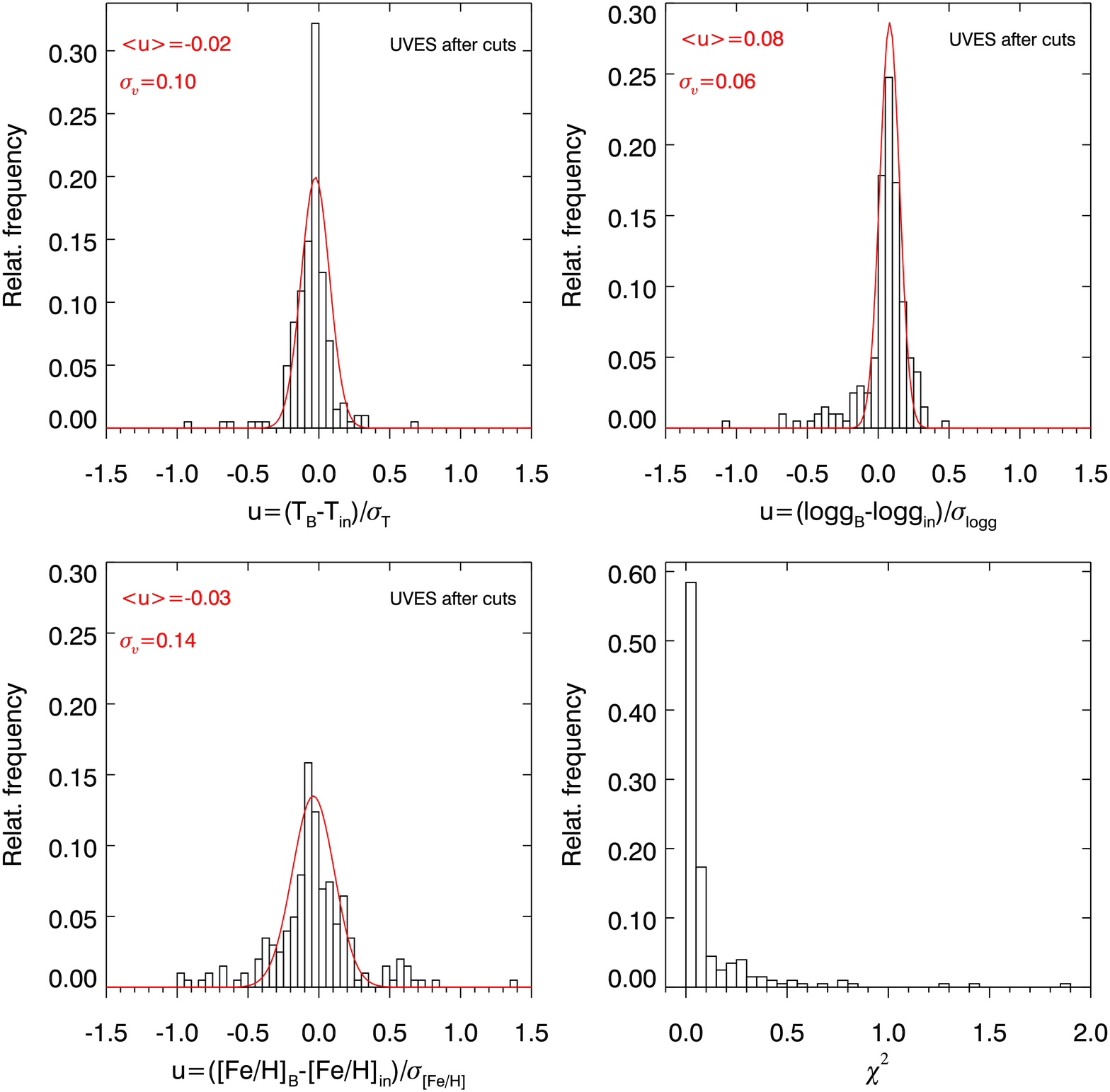}
\caption{Top 4 panels: The stellar parameters recovered by BeSPP for the ideal
synthetic stellar sample drawn from the tracks (see text).  Bottom 4 panels: The
stellar parameters ($\teff$, $\logg$,  $\feh$) recovered by BeSPP compared with
the input spectroscopic values for the UVES stars. The total $\chi^2$ was
computed summing over the three spectroscopic parameters and using only the
observational  errors.}
\label{mc}
\end{figure}

Figure~\ref{mc} (top panels) shows the results for the stars in the synthetic
samples that fulfil the age uncertainty criteria. The top and  bottom panels
correspond to $4.0 \leq \logg \leq 4.6$ and $3.5  \leq  \logg  <  4.0$,
respectively.  The quantity  $u_i$ is our measure of  bias and dispersion, and
it is defined  as $u_i= (X_{{\rm B,}i}  - X_{{\rm in,}i})/\sigma_{{\rm X},i}$,
where $X_i$ denotes  the mass or age of a synthetic  star $i$ and $\sigma_{{\rm 
 X,}i}$ the uncertainty returned by BeSPP. The Gaussian fit to the distribution
is overplotted in red. The average difference and dispersion are also quoted.
We find that for subgiant  stars, $3.5  \leq  \logg  <  4.0$, systematics are 
virtually absent and the dispersion is well  below unity, a result that probably
derives from the faster evolution of stars in this phase. In the range  $4.0
\leq \logg  \leq 4.6$  the method  tends to overestimate masses and to
underestimate ages by $44\%$ and $36\%$ of the mean uncertainty, respectively. 
The dispersion is very close to unity, consistent with the expected distribution
of errors. The systematic uncertainties introduced by BeSPP in  the
determination of the age of dwarf stars will be typically 1.3~Gyr for a 12~Gyr
old star, down to $\sim 0.5$~Gyr for a 5~Gyr  star.

One may also question whether there are systematic differences in the parameter
scales between theory and observations. For example, the treatment of convection
and surface boundary effects in stellar evolution models may lead to a
difference in the $\teff$ scales compared to spectroscopic predictions.
Likewise, systematic uncertainties in stellar parameters may cause serious
offsets in surface vs stellar evolution scales. To test this, we compared the
stellar parameters ($\teff$, $\logg$,  $\feh$) recovered by BeSPP  with the
input spectroscopic values (Fig.~\ref{mc}, bottom panel). The left-bottom panel
shows the total $\chi^2$ computed summing over the three spectroscopic
parameters and using only the observational errors. The agreement is excellent,
with dispersion values $\sigma$ of the order of $0.1$, ruling out the
possibility of a systematic mismatch between the observed data and stellar
models.

Another  method to  validate the  accuracy of the Bayesian results is to compare
them with that obtained  by independent methods, e.g. from astero-seismology.
The number of stars observed in iDR1 is too small to draw rigorous conclusions.
While  there is no  independent  information for  the majority of  the UVES 
stars, we  can use the library  of FGK Benchmark stars, which  was built for the
Gaia-ESO survey (\citealt{2013arXiv1309.1099J}, \citealt{2014arXiv1403.3090B},
Heiter et al. in prep.). We have compared the masses from BeSPP for the
reference parameters with independent values derived from astero-seismic
analysis (available for 12 stars). This comparison revealed a mean offset
between BeSPP and astero-seismology of $0.07 \pm  0.08 \Msun$. The agreement is
satisfactory, and it confirms the accuracy of the algorithm that we use for the
age and mass calculations.

\section{Star formation rate}

We follow the recipe of Schoenrich \& Bergemann (2013), where the SFR is assumed
proportional to 
\begin{equation}
P(\tau,\feh)= \left\{ 
\begin{array}{ccc} 
0 & \rm{if} & \tau > 15 \rm{Gyr} \\ 
1 & \rm{if} & 11 \rm {Gyr} \leq \tau \leq 15 \rm{Gyr} \\ 
\exp{\left(\frac{\tau - 11{\rm Gyr}}{\sigma_\tau}  \right)} & \rm{if} & \tau <
11 \rm {Gyr} 
  \end{array}
\right. 
\end{equation}
where
\begin{eqnarray}
\sigma_\tau= \left\{ 
\begin{array}{ccc} 
1.5 \, \rm{Gyr} & \rm{if} & \feh < -0.9 \\
1.5 + 7.5 \frac{\feh+0.9}{0.4}\, \rm{Gyr} & \rm{if} & -0.9 \leq \feh \leq -0.5
\\
9 \, \rm{Gyr} & \rm{if} & -0.5 < \feh.
\end{array}
\right. \nonumber
\end{eqnarray}

\end{appendix}
\bibliographystyle{aa}
\end{document}